\documentclass[12pt]{article}

\usepackage{a4wide,epsfig,psfrag,cite}
\usepackage{amsmath,amssymb,scalefnt}
\usepackage{color}
\usepackage[font=small]{caption}

\graphicspath{ {figs/} }

\definecolor{darkgreen}{cmyk}{1.0,0,1.0,0.61}

\definecolor{light-gray}{gray}{0.95}

\newcommand{\asfivepi}{\bigg(\!\frac{\alpha_s^{(5)}(M_h)}{\pi}\!\bigg)}
\newcommand{\acone}{a_1}
\newcommand{\acfour}{a_4}
\newcommand{\acfive}{a_5}
\newcommand{\pitwo}{\pi^2}
\newcommand{\pifour}{\pi^4}
\newcommand{\zetathree}{\zeta_3}
\newcommand{\zetafive}{\zeta_5}
\newcommand{\IPi}{i\pi}
\newcommand{\lHt}{l_{Ht}}

\allowdisplaybreaks

\begin{document}

\title{\vskip-3cm{\baselineskip14pt
    \begin{flushleft}
      \normalsize FERMILAB-PUB-21-213-T\\
      TTP21-010\\
      P3H-21-028\\
  \end{flushleft}}
  \vskip1.5cm
  Higgs boson decay into photons at four loops
}

\author{
  Joshua Davies$^{\, 1}$ and Florian Herren$^{\, 2}\footnote{Address until 18.01.2021: {\it Institut f{\"u}r Theoretische Teilchenphysik, Karlsruhe Institute of Technology (KIT), 76128 Karlsruhe, Germany}}$
  \\[1em]
  {\small\it $^{1}$ Department of Physics and Astronomy, University of Sussex}\\
  {\small\it Brighton, BN1 9HQ, UK}\\
  {\small\it $^{2}$ Fermi  National  Accelerator  Laboratory}\\
  {\small\it Batavia,  IL,  60510,  USA}
}

\date{}

\maketitle

\begin{abstract}
Future precision measurements of Higgs boson decays will determine the branching fraction for the decay into two photons with a precision at the one percent level.
To fully exploit such measurements, equally precise theoretical predictions need to be available.
To this end we compute four-loop QCD corrections in the large top quark mass expansion to the Higgs boson--photon form factor, which enter the two-photon decay width at next-to-next-to-next-to-leading order.
Furthermore we obtain corrections to the two-photon decay width stemming from the emission of additional gluons,
which contribute for the first time at next-to-next-to-leading order.
Finally, we combine our results with other available perturbative corrections and estimate the residual uncertainty due to missing higher-order contributions.
\end{abstract}

\thispagestyle{empty}

\newpage

\section{Introduction}
The decay of the Higgs boson into two photons was among the main discovery channels at the Large Hadron Collider (LHC) \cite{Aad:2012tfa,Chatrchyan:2012ufa} and plays a crucial role in precision studies of its properties.
Among these are measurements of its mass, see e.g. \cite{Sirunyan:2020xwk}, as well as the study of the interference between the di-photon signal with the continuum background, which leads to a shift in the di-photon mass spectrum
and in production rates, thus allowing for studies of the total Higgs boson width \cite{Dixon:2003yb,Martin:2012xc,Martin:2013ula,Dixon:2013haa,Coradeschi:2015tna,Campbell:2017rke}.
In addition, future colliders will allow the measurement of ratios of branching fractions at the sub-percent level (see e.g. \cite{Benedikt:2018csr}), thus demanding
theoretical predictions of partial decay widths at the same level of precision.

While the next-to-next-to-leading order (NNLO) top quark--induced corrections amount to less than one percent of the full decay width,
it is possible that their smallness is accidental and higher order corrections are of the same size.
Furthermore, the scale and scheme dependence need to be quantified for a serious estimate of the theoretical uncertainty.
This work tries to cover both of these points by performing a four-loop computation of the top quark--induced QCD corrections, combining them with all available higher-order corrections and finally
quantifying the theoretical uncertainty on the partial decay width into photons.

The amplitude for this decay can be written as
\begin{align}
\mathcal{A}_{h\rightarrow\gamma\gamma} = \epsilon_1^\mu\epsilon_2^\nu\left(g_{\mu\nu}(p_1\cdot p_2) - p_{2\mu}p_{1\nu}\right)\,A(s),
\end{align}
where $p_i$ and $\epsilon_i$ are the momenta and polarization vectors of the photons, with $p_i^2 = 0$ and $\epsilon_i\cdot p_i = 0$. The Lorentz-scalar function $A(s)$ only depends on
the centre-of-mass energy $s = 2 p_1\cdot p_2$, as well as the masses of internal particles. It enters the $h\rightarrow\gamma\gamma$ decay width as
\begin{align}
\Gamma_{h\rightarrow\gamma\gamma} = \frac{M_h^3}{64\pi}\left|A(s)\right|^2~,
\end{align}
where $M_h$ is the mass of the Higgs boson. The leading-order (LO) (one-loop) contributions to $A(s)$ have been known for a long time \cite{Ellis:1975ap,Shifman:1979eb}. They can be decomposed into W-Boson contributions, $A_W$,
as well as fermionic contributions, $A_{t,b,\tau}$, due to the top quark, bottom quark and tau lepton.

Next-to-leading-order (NLO) QCD corrections to $A_t$ have been computed in the limit of a large top quark mass \cite{Zheng:1990qa,Djouadi:1990aj,Dawson:1992cy,Melnikov:1993tj,Inoue:1994jq}
and later with the full mass dependence \cite{Spira:1995rr,Fleischer:2004vb,Harlander:2005rq,Aglietti:2006tp}, thus also providing NLO QCD corrections to $A_b$.
NNLO QCD corrections
in an expansion for a large top quark mass have been obtained in \cite{Steinhauser:1996wy,Maierhofer:2012vv} and recently, exact numerical results became available \cite{Niggetiedt:2020sbf}.
Partial results in the limit of an
infinitely heavy top quark are known at next-to-next-to-next-to-leading order (N$^3$LO) \cite{Sturm:2014nva}.
NLO electroweak corrections, for which the above decomposition is not possible, have been obtained in \cite{Fugel:2004ug,Degrassi:2005mc,Passarino:2007fp,Actis:2008ts}.

Virtual QCD corrections to $A_{t,b}$ do not lead to infrared singularities, so one does not need
to include real-radiation contributions to obtain a finite result. Nonetheless, these real-radiation
contributions should be considered since in principle the emission of additional,
possibly soft, gluons leads to a shift in the di-photon mass spectrum. At NLO, such
contributions vanish due to colour conservation; the first non-vanishing contribution arises from the
process $h \rightarrow \gamma\gamma gg$ at NNLO. To our knowledge, the size of these contributions has not been
investigated in the literature.

In this work, we concentrate on $A_t$ and compute the N$^3$LO virtual contributions in an
asymptotic expansion for a large top quark mass.
We additionally consider the NNLO real-radiation contributions in the same approximation.
The rest of this article is structured as follows:
in Section \ref{sec:virt} we present the method by which we compute the N$^3$LO virtual corrections,
followed by a discussion of the NNLO real-radiation contributions in Section \ref{sec:real}.
Finally we discuss the numerical impact of both of these contributions in Section \ref{sec:res}.


\section{\label{sec:virt}Calculation of N$^3$LO virtual corrections}
Our computation of the virtual corrections follows \cite{Davies:2019wmk}, in which we computed the top quark contributions to the gluon-gluon-Higgs form factor. In the following, we discuss
the computational setup and present analytical results for the $h\rightarrow\gamma\gamma$ form factor. These corrections can not be obtained from the results of \cite{Davies:2019wmk},
since starting from three loops it is not possible to simply substitute colour factors.

\subsection{Computational setup}
We generate 2 one-loop, 12 two-loop, 206 three-loop and 5062 four-loop diagrams contributing to $A_t$ using {\tt QGRAF} \cite{Nogueira:1991ex}. We then generate {\tt FORM} \cite{Ruijl:2017dtg} code for
each diagram using {\tt q2e} \cite{Seidensticker:1999bb,Harlander:1997zb}, compute the colour factors using {\tt COLOR} \cite{vanRitbergen:1998pn} and perform an expansion-by-subgraph \cite{Smirnov:2012gma}
in the limit $M_t^2 \gg s = M_h^2$ using {\tt exp} \cite{Seidensticker:1999bb,Harlander:1997zb}.
As a result, all diagrams are mapped onto one- to four-loop massive tadpole integrals and one- to three-loop massless form factor integrals, separating the two scales $M_t$ and $s$.

Both types of integrals have received a lot of attention in the literature; all required master integrals are known analytically \cite{Laporta:2002pg,Schroder:2005va,Chetyrkin:2006dh,Lee:2010hs,Baikov:2009bg,Heinrich:2009be,
Lee:2010ik,Gehrmann:2010ue,Gehrmann:2010tu}. Due to the asymptotic expansion, we need to deal with tensor integrals, i.e. integrals for which we can not re-write all numerator structures
in terms of inverse propagators of the integral family. As a consequence, we have to perform a tensor reduction of the numerator structures. Since in the fully hard region of the asymptotic expansion only tadpole
integrals can appear, and in all other regions both types of integrals appear, it is advantageous to perform the tensor reduction for the tadpole integrals. For one- and two-loop integral families, there are general
algorithms for treating tensor tadpole integrals of an arbitrary rank \cite{Chetyrkin:1993rv}. These are implemented in {\tt MATAD} \cite{Steinhauser:2000ry}.
At three and four loops, we have implemented routines to reduce tensors up to rank 10. This is a sufficient
rank to expand the form factor up to $M_t^{-6}$.
After tensor reduction we perform an integration-by-parts reduction of the three and four loop tadpole integrals,
as well as the massless form factor integrals, in order to obtain the $h\rightarrow\gamma\gamma$ form factor in
terms of master integrals. For this we use {\tt LiteRed} \cite{Lee:2012cn,Lee:2013mka} and {\tt FIRE6}\cite{Smirnov:2019qkx}.

We then proceed to renormalize the bare strong coupling constant $\alpha_s^0$ and the top quark mass $m_t^0$ in the $\overline{\mathrm{MS}}$ scheme. For this we require the strong coupling renormalization constant at two loops,
(since the leading-order contribution to the form factor does not depend on $\alpha_s$) and the quark mass renormalization constant at three loops\footnote{Both renormalization constants have been known for a long time. We take their
expressions from \cite{Chetyrkin:2017bjc}, where they are expressed in terms of $\mathrm{SU}(N)$ colour factors and are available as computer-readable files.}.
This renders the form factor finite, as there are no infrared divergences
present. This is in contrast to the gluon-gluon-Higgs form factor discussed in Ref.~\cite{Davies:2019wmk}. Since the top quark does not appear as a dynamical degree of freedom at energy scales of the order of the Higgs boson mass, we
transform $\alpha_s$ from the six- to the five-flavour scheme, using the two-loop decoupling constant\footnote{Since we express our results in terms of general $\mathrm{SU}(N)$ colour factors, we take the expressions from
\cite{Gerlach:2018hen}, where they are available in a computer-readable form.}.
Furthermore, we also produce the form factor with the top quark mass renormalized in the on-shell scheme, which we obtain
from the $\overline{\mathrm{MS}}$ result by applying the $\overline{\mathrm{MS}}$-OS relation at three loops \cite{Chetyrkin:1999ys,Chetyrkin:1999qi,Melnikov:2000qh,Marquard:2007uj}.

\subsection{Analytical results}
In the following we present analytical results for the $h\rightarrow\gamma\gamma$ form factor at four loops.
To this end, we decompose the form factor into three contributions,
\begin{align}
\label{eq::at}
A_t &= \hat{A}_t\left(Q_t^2 A_{t,t} + Q_t \sum_{f} Q_f A_{t,f} + \sum_f Q_f^2 A_{f,f}\right)~.
\end{align}
Here, $Q_t = \frac{2}{3}$ is the electric charge of the top quark and the sum over $f$ runs over all five massless quark flavours ($u,d,s,c,b$), where $Q_u = Q_c = \frac{2}{3}$ and $Q_d = Q_s = Q_b = -\frac{1}{3}$.
The sums over the quark charges in Eq.~(\ref{eq::at}) evaluate to
\begin{align}
\sum_{f} Q_f = \frac{1}{3}\quad\text{and}\quad\sum_{f} Q_f^2 = \frac{11}{9}~
\end{align}
and the overall prefactor is given by
\begin{align}
\hat{A}_t &= \frac{2\alpha}{3\pi}N_c \sqrt{\sqrt{2}G_F}~,
\end{align}
where $\alpha$ is the fine structure constant, $N_c = 3$ the number of colours and $G_F$ is Fermi's constant.

\begin{figure}[h]
\centering
\includegraphics[width=\textwidth]{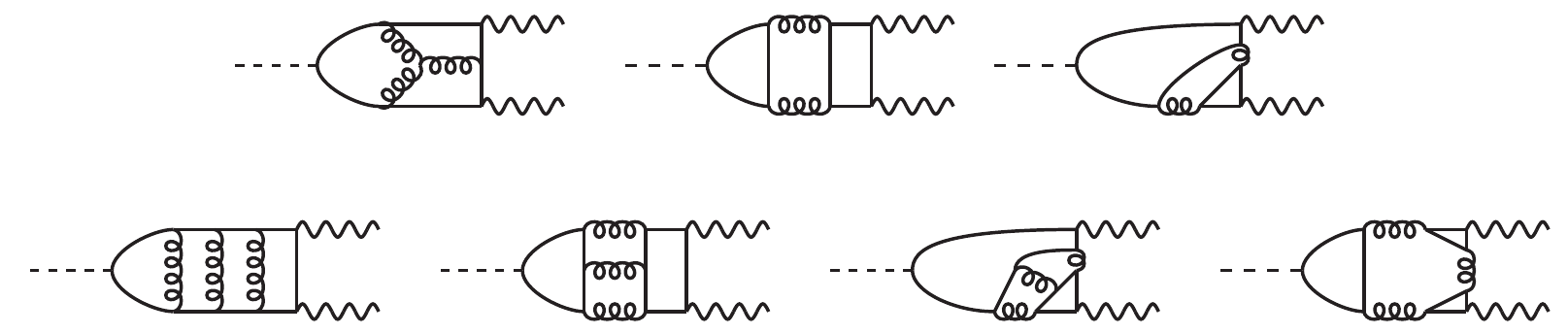}
\caption[Virtual corrections]{\label{fig::furry}Sample Feynman diagrams contributing to the $h\rightarrow\gamma\gamma$ form factor at three and four loops.
Dashed, solid, wavy and curly lines denote Higgs bosons, quarks, photons and gluons respectively.
The right-most diagrams in both lines sum to zero due to Furry's theorem\cite{Furry:1937zz}.}
\end{figure}

Sample Feynman diagrams which contribute to $A_t$ at three- and four-loop order are shown in
Fig.~\ref{fig::furry}.
Note that at three loops, diagrams such as the last diagram in the first line, which contribute to $A_{t,f}$, sum to zero due to Furry's theorem \cite{Furry:1937zz}.
Furthermore, starting from four loops, there are diagrams such as the last of the second line, in which the photons couple to different light-quark loops; these also sum to zero for the same reason.

For the three contributions of Eq.~(\ref{eq::at}) we obtain, for $\mu^2 = M_h^2$ in the $OS$ scheme,
{\footnotesize
\begin{align}
A_{t,t} &=\:
%
       1
       + \frac{7}{120}
          \rho
       + \frac{1}{168}
          \rho^2
       + \frac{13}{16800}
          \rho^3
%
    + \asfivepi \bigg(
       - 1
       + \frac{61}{270}
          \rho
       + \frac{554}{14175}
          \rho^2
       + \frac{104593}{15876000}
          \rho^3
    \bigg)
\nonumber\\&
%
    + \asfivepi^2 \bigg(
       - \frac{7}{6}
       - \frac{23}{12} \lHt
       + \rho   \bigg[
          - \frac{4904561}{622080}
          + \frac{7}{1080} \pitwo
          + \frac{7}{540} \pitwo \acone
          + \frac{206951}{27648} \zetathree
          + \frac{1403}{3240} \lHt
          \bigg]
\nonumber\\&
       + \rho^2   \bigg[
          - \frac{14134687057}{10450944000}
          + \frac{1}{756} \pitwo
          + \frac{1}{378} \pitwo \acone
          + \frac{18180533}{13271040} \zetathree
          + \frac{1}{360} \IPi
          + \frac{24539}{340200} \lHt
          \bigg]
\nonumber\\&
       + \rho^3   \bigg[
          - \frac{25502795129137657}{1966449623040000}
          + \frac{13}{50400} \pitwo
          + \frac{13}{25200} \pitwo \acone
          + \frac{28639831757}{2642411520} \zetathree
          + \frac{41}{129600} \IPi
\nonumber\\&
          + \frac{2345369}{190512000} \lHt
          \bigg]
    \bigg)
%
\nonumber\\&
    + \asfivepi^3 \bigg(
       - \frac{21467}{864}
       + \frac{6425}{288} \zetathree
       - \frac{62}{9} \lHt
       - \frac{529}{144} \lHt^2
       + \rho   \bigg[
          - \frac{235989087607}{7390310400}
          + \frac{2800187}{2332800} \pitwo
\nonumber\\&
          - \frac{895}{1944} \pitwo \acone
          + \frac{26268927251}{547430400} \zetathree
          - \frac{91591991}{26127360} \acone^4
          + \frac{90293687}{26127360} \pitwo \acone^2
          + \frac{1512779543}{3135283200} \pifour
          - \frac{91591991}{1088640} \acfour
\nonumber\\&
          - \frac{46}{2835} \acone^5
          + \frac{46}{1701} \pitwo \acone^3
          - \frac{10073}{25920} \zetathree \pitwo
          + \frac{19133}{136080} \pifour \acone
          - \frac{839243}{18144} \zetafive
          + \frac{368}{189} \acfive
          - \bigg\{ \frac{22153403}{746496} - \frac{161}{6480} \pitwo
\nonumber\\&
              - \frac{161}{3240} \pitwo \acone - \frac{4759873}{165888} \zetathree \bigg\} \lHt
          + \frac{32269}{38880} \lHt^2
          \bigg]
       + \rho^2   \bigg[
          - \frac{2222912377084634113}{67789839237120000}
          + \frac{1985747}{8164800} \pitwo
\nonumber\\&
          - \frac{44311}{510300} \pitwo \acone
          + \frac{9990549815126663}{334764638208000} \zetathree
          - \frac{199893333511}{112086374400} \acone^4
          + \frac{28393807873}{16012339200} \pitwo \acone^2
\nonumber\\&
          + \frac{26390124812159}{107602919424000} \pifour
          - \frac{199893333511}{4670265600} \acfour
          + \frac{32309}{277992} \acone^5
          - \frac{161545}{833976} \pitwo \acone^3
          - \frac{1439}{18144} \zetathree \pitwo
\nonumber\\&
          - \frac{5014747}{33359040} \pifour \acone
          + \frac{10527619}{1482624} \zetafive
          - \frac{161545}{11583} \acfive
          + \frac{1963}{29160} \IPi
          - \bigg\{ \frac{323396467271}{62705664000} - \frac{23}{4536} \pitwo
             - \frac{23}{2268} \pitwo \acone
\nonumber\\&
             - \frac{418152259}{79626240} \zetathree \bigg\} \lHt
          + \frac{564397}{4082400} \lHt^2
          \bigg]
       + \rho^3   \bigg[
          - \frac{8299889702618880183134701759}{45319709343993141657600000}
          + \frac{36852569}{762048000} \pitwo
\nonumber\\&
          - \frac{758599}{47628000} \pitwo \acone
          + \frac{5920387676337701221931}{6394315251357057024000} \zetathree
          - \frac{260319544831343}{36436967424000} \acone^4
          + \frac{1821732431724041}{255058771968000} \pitwo \acone^2
\nonumber\\&
          + \frac{2010057846448848593}{642748105359360000} \pifour
          - \frac{260319544831343}{1518206976000} \acfour
          + \frac{151648997}{1459458000} \acone^5
          - \frac{151648997}{875674800} \pitwo \acone^3
\nonumber\\&
          - \frac{18707}{1209600} \zetathree \pitwo
          + \frac{1097561849}{70053984000} \pifour \acone
          - \frac{179329653863}{3113510400} \zetafive
          - \frac{151648997}{12162150} \acfive
          + \frac{1455731}{195955200} \IPi
\nonumber\\&
          - \bigg\{ \frac{586464088779883871}{11798697738240000} - \frac{299}{302400} \pitwo - \frac{299}{151200} \pitwo \acone
              - \frac{658716130411}{15854469120} \zetathree \bigg\} \lHt
          + \frac{53943487}{2286144000} \lHt^2
          \bigg]
    \bigg)\,,
\nonumber\\
\\
A_{t,f} &=
%
%
%
%
    \asfivepi^3 \bigg(
       \frac{55}{108}
       - \frac{10}{9} \zetathree
       + \rho   \bigg[
          \frac{3545}{10368}
          + \frac{245}{648} \zetathree
          - \frac{17}{1944} \pifour
          \bigg]
       + \rho^2   \bigg[
          \frac{9222149}{100776960}
          + \frac{2129}{15552} \zetathree
\nonumber\\&
          - \frac{3137}{1166400} \pifour
          \bigg]
       + \rho^3   \bigg[
          \frac{42425653493}{1128701952000}
          + \frac{161441}{583200} \zetathree
          - \frac{207383}{54432000} \pifour
          \bigg]
    \bigg)\,,
\\
\nonumber\\
A_{f,f} &=
%
%
%
    \asfivepi^2 \bigg(
          - \frac{13}{12}
          + \frac{2}{3} \zetathree
          - \frac{1}{6} \IPi
          + \frac{1}{6} \lHt
       + \rho   \bigg[
          - \frac{3493}{194400}
          + \frac{7}{180} \zetathree
          - \frac{19}{6480} \IPi
          + \frac{19}{6480} \lHt
          \bigg]
\nonumber\\&
       + \rho^2   \bigg[
          - \frac{3953}{6350400}
          + \frac{1}{252} \zetathree
          + \frac{1}{60480} \IPi
          - \frac{1}{60480} \lHt
          \bigg]
\nonumber\\&
       + \rho^3   \bigg[
          - \frac{3668899}{171460800000}
          + \frac{13}{25200} \zetathree
          + \frac{53}{2126250} \IPi
          - \frac{53}{2126250} \lHt
          \bigg]
    \bigg)
%
\nonumber\\&
   + \asfivepi^3 \bigg(
          - \frac{10337}{648}
          + \frac{529}{1296} \pitwo
          + \frac{27}{2} \zetathree
          + \frac{23}{9720} \pifour
          - \frac{25}{9} \zetafive
          - \bigg\{ \frac{167}{36} - \frac{23}{9} \zetathree \bigg\} \IPi
          + \frac{35}{72} \lHt
\nonumber\\&
          + \frac{23}{72} \lHt^2
       + \rho   \bigg[
          - \frac{19038301}{26244000}
          + \frac{35131}{1399680} \pitwo
          + \frac{424577}{622080} \zetathree
          + \frac{161}{1166400} \pifour
          - \frac{35}{216} \zetafive
          - \bigg\{ \frac{1887643}{6998400}
\nonumber\\&
              - \frac{161}{1080} \zetathree \bigg\} \IPi
          + \frac{1405609}{6998400} \lHt
          + \frac{1441}{58320} \lHt \IPi
          - \frac{1571}{233280} \lHt^2
          \bigg]
       + \rho^2   \bigg[
          \frac{328389003179}{512096256000}
          + \frac{1709}{653184} \pitwo
\nonumber\\&
          - \frac{352361479}{696729600} \zetathree
          + \frac{23}{1632960} \pifour
          - \frac{25}{1512} \zetafive
          - \bigg\{ \frac{132673099}{5486745600} - \frac{23}{1512} \zetathree \bigg\} \IPi
          + \frac{119580763}{5486745600} \lHt
\nonumber\\&
          + \frac{31903}{8709120} \lHt \IPi
          - \frac{6491}{3483648} \lHt^2
          \bigg]
       + \rho^3   \bigg[
          \frac{3187611042300355367}{2488787804160000000}
          + \frac{7228849}{20995200000} \pitwo
\nonumber\\&
          - \frac{24814688746213}{23410114560000} \zetathree
          + \frac{299}{163296000} \pifour
          - \frac{13}{6048} \zetafive
          - \bigg\{ \frac{339024973909}{123451776000000} - \frac{299}{151200} \zetathree \bigg\} \IPi
\nonumber\\&
          + \frac{46985544667}{17635968000000} \lHt
          + \frac{54757123}{97977600000} \lHt \IPi
          - \frac{64119043}{195955200000} \lHt^2
          \bigg]
    \bigg)\,.
\end{align}
}%
Here, $\zeta_n$ is the Riemann $\zeta$-function evaluated at $n$, $a_n = \mathrm{Li}_n(1/2)$, $\rho = M_h^2/M_t^2$
and $l_{Ht} = \log(M_h^2/M_t^2)$.
To reduce the size of the expressions the colour factors have been set to the following values:
\begin{align}
C_A = 3~,\quad C_F = \frac{4}{3}~,\quad T_F = \frac{1}{2}\quad \text{and}\quad d_F^{abc}d_F^{abc} = \frac{5}{6}~,
\end{align}
however the full expressions, including the colour factors and for an arbitrary scale $\mu^2$, are provided in the ancillary files of this paper.

This result passes a number of cross-checks. Firstly our result is finite after renormalization and, in the abelian limit, agrees with \cite{Davies:2019wmk}. Secondly, the part stemming from the fully hard region in the asymptotic expansion agrees with \cite{Sturm:2014nva}.

\section{\label{sec:real}Calculation of NNLO real corrections}
Starting from NNLO in QCD, corrections with two gluons in the final state can also be taken into account\footnote{Contributions with one final-state gluon arise at NLO, which vanish due to colour conservation.}.
While these corrections do not lead to infrared divergences and do not lead to large logarithms, they can potentially impact the di-photon invariant mass spectrum, thus possibly influencing extractions of the total width and Higgs
boson mass.
In the following we compute the contributions to the inclusive Higgs boson decay width.

\begin{figure}[h]
\centering
\includegraphics[width=0.7\textwidth]{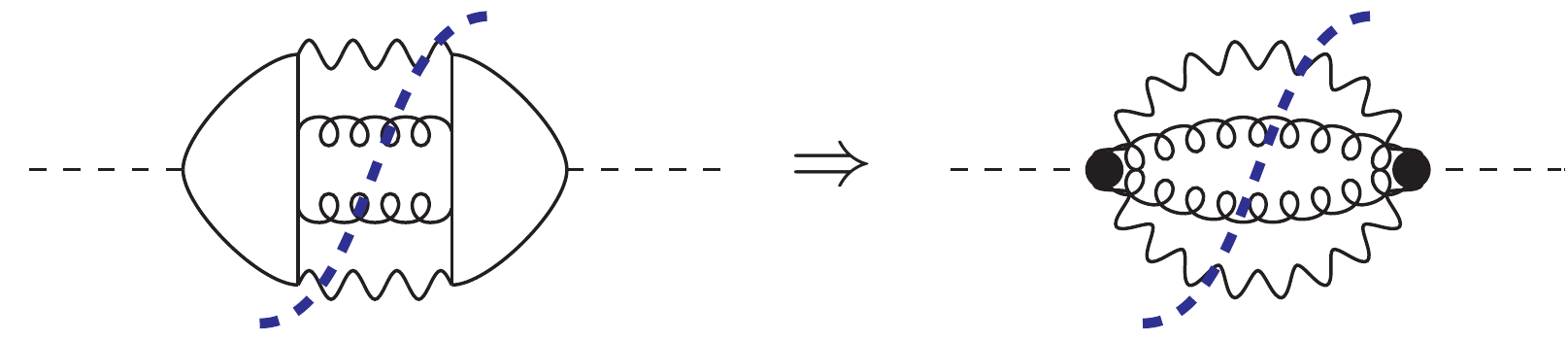}
\caption[Real corrections]{\label{fig::real} The building-block approach for the NNLO real corrections. Each building block starts at $M_t^{-4}$ and incorporates the contributions of 24 individual diagrams.
The dashed line denotes the cut through the final-state photon and gluon propagators.}
\end{figure}

We employ the method of reverse unitarity \cite{Anastasiou:2002yz} to express the phase-space integral over the squared decay amplitude for $h\rightarrow\gamma\gamma gg$ as
cut two-point forward-scattering diagrams. In each of these diagrams the Higgs bosons couple to different top-quark loops, which are connected to each other by the cut gluons and photons.
A sample diagram is shown in Fig.~\ref{fig::real}. A direct large-mass expansion of such five-loop diagrams is computationally expensive, thus we adopt the building-block approach
described in \cite{Davies:2019xzc,Davies:2019esq}; we pre-compute the 24 one-loop diagrams contributing to the $h\gamma\gamma gg$ vertex in the large-mass expansion. Since there are only hard
subgraphs, the result is polynomial in all external momenta and can be used as an effective vertex. As a consequence, instead of 120 five-loop diagrams, we only expand 24 one-loop diagrams which then are
inserted into the cut three-loop two-point diagram shown in Fig.~\ref{fig::real}. The resulting diagram can be computed using {\tt MINCER} \cite{Larin:1991fz}.

While each of the one-loop five-point diagrams starts at $M_t^0$ in the large-mass expansion, their sum starts at $M_t^{-4}$.
Consequently, the expansion of each five-loop diagram also starts at $M_t^0$ but their sum starts at
$M_t^{-8}$. This can be understood by the fact that gauge-invariant operators leading to a $\gamma\gamma gg$ interaction need to contain two gluonic and two photonic field-strength tensors and thus are dimension-8 operators.

Computing the individual cut five-loop diagrams would require the computation of five terms in the large-mass expansion, the first four of which ultimately cancel in the sum.
Using instead the building-block approach allows us to compute contributions to $\Gamma_{h\rightarrow \gamma\gamma g g}$ to order $M_t^{-12}$ without much computational effort.

We obtain
\begin{align}
\Gamma_{h\rightarrow \gamma\gamma g g} &= \frac{M_h^3}{64\pi}\hat{A}_t^2\left(\frac{\alpha_s^{(5)}}{\pi}\right)^2\left(\frac{17}{34020000}\rho^4 + \frac{37}{136080000}\rho^5 + \frac{219749}{2240421120000}\rho^6\right)~.
\end{align}

We note that by using {\tt FORCER}\cite{Ruijl:2017cxj} to compute four-loop two-point phase-space integrals, it would
be a relatively straightforward task to compute also the N$^3$LO real and real-virtual contributions.
As demonstrated later in Eq.~(\ref{eq::nnlo-real-numerics}) the numerical impact of the NNLO contributions
is very small, so we do not perform this computation here.

\section{\label{sec:res}Numerical results}
In the following we study the numerical impact of the N$^3$LO top quark contributions, including renormalization scale and scheme dependence, before
combining them with the NNLO QCD corrections for bottom and charm quark loops and the NLO electroweak corrections. Finally, we discuss sources of uncertainty
on the partial decay width.

\subsection{\label{sec::toponly} N$^3$LO top quark contributions}
In the following we discuss the numerical impact of the corrections obtained in the previous sections on the Higgs boson decay width into photons.
To this end, we employ the exact LO result for $A(s)$, neglecting light fermion loops, and combine it with the large-mass expansion results for $A_t$ starting from NLO.

The parameters entering the evaluation are given by \cite{Zyla:2020zbs}:
\begin{align}
\alpha &= 1/137.036~,\quad G_F = 1.1663787 \times 10^{-5}~\text{GeV}^{-2},\quad M_W = 80.379~\text{GeV},\nonumber\\
\alpha^{(5)}_s\left(M_Z\right) &= 0.1179~,\quad M_Z = 91.1876~\text{GeV},\quad M_t = 172.76~\text{GeV},\quad M_h = 125.10~\text{GeV}.\label{eq::input1}
\end{align}
We use the program \texttt{RunDec} \cite{Chetyrkin:2000yt,Herren:2017osy} to evolve the strong coupling constant to different renormalization scales and to obtain the top quark mass in the $\overline{\mathrm{MS}}$ scheme.

For the renormalization scale $\mu = M_h$ we obtain the decay width with the top quark mass renormalized in both, the on-shell and the $\overline{\mathrm{MS}}$ scheme
\begin{align}
\label{eq::Gamma-muMh-MtOS}
\Gamma_{h\rightarrow \gamma\gamma}^{\text{OS}} \times 10^{6} \text{GeV}^{-1} &= 9.1322 + 0.1558 + 0.0029 - 0.0031 = 9.2878~,\\
\Gamma_{h\rightarrow \gamma\gamma}^{\overline{\text{MS}}} \times 10^{6} \text{GeV}^{-1} &= 9.1188 + 0.1639 + 0.0070 - 0.0023 = 9.2874~,
\end{align}
where we have separated the LO, NLO, NNLO and N$^3$LO contributions.
While the NNLO corrections are two orders of magnitude smaller than the NLO corrections,
the N$^3$LO corrections are larger than the NNLO corrections in the on-shell scheme, but negative.
In the $\overline{\text{MS}}$ scheme they are also negative and the same order of magnitude, partially
cancelling the effect of the NNLO corrections.
In both cases, the effect on the decay width is small, resulting in a $0.034\%$ and a $0.024\%$
decrease, respectively.

\begin{figure}
\centering
\includegraphics{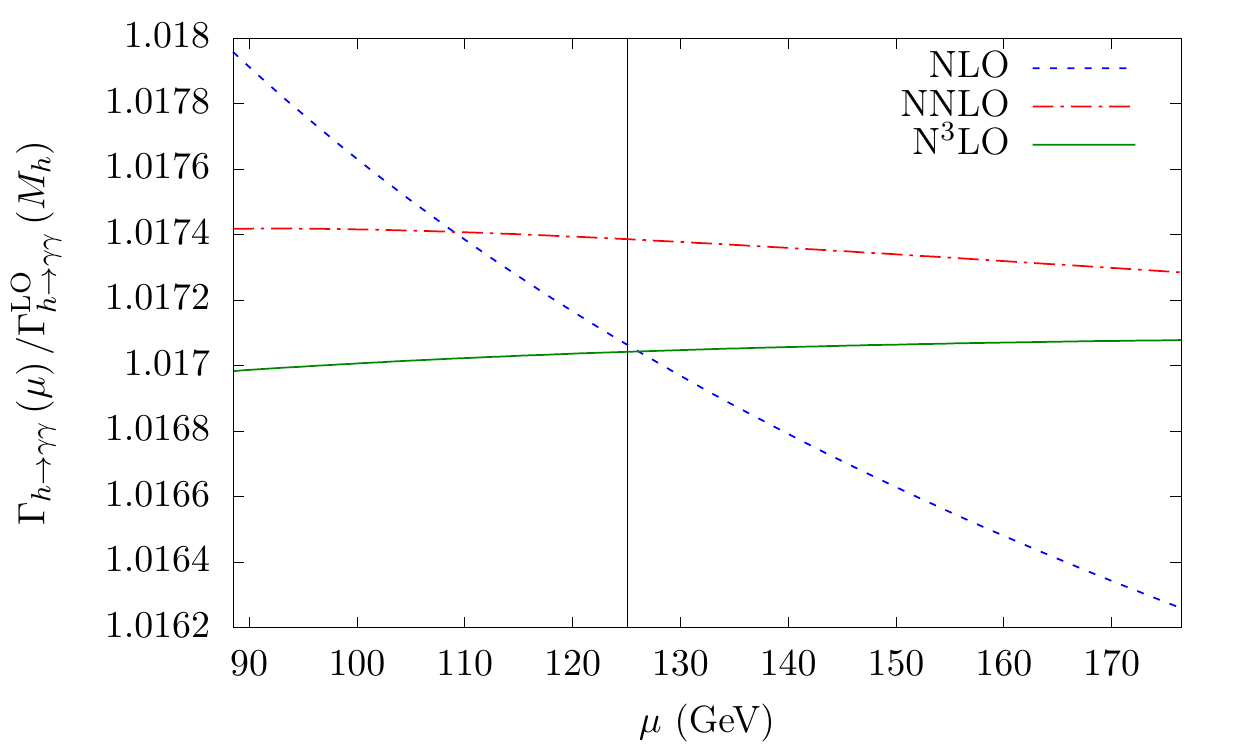}
\caption{\label{plt::os}
The perturbative expansion of the partial decay width, normalized to
$\Gamma_{h\rightarrow \gamma\gamma}^{\mathrm{LO}}(M_h)$.
The top quark mass is renormalized
in the on-shell scheme. The vertical line denotes $\mu = M_h$.}
\end{figure}

\begin{figure}
\centering
\includegraphics{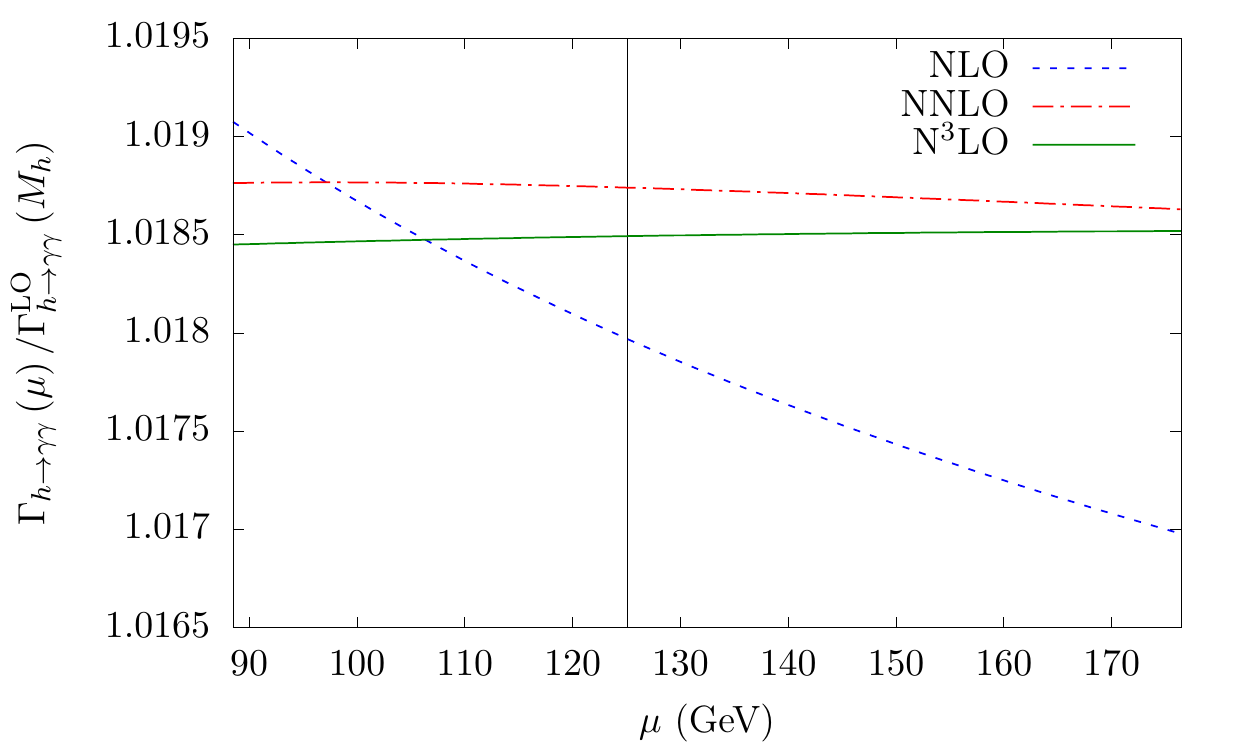}
\caption{\label{plt::ms}
The perturbative expansion of the partial decay width, normalized to
$\Gamma_{h\rightarrow \gamma\gamma}^{\mathrm{LO}}(M_h)$.
The top quark mass is renormalized
in the $\overline{\text{MS}}$ scheme. The vertical line denotes $\mu = M_h$.}
\end{figure}

This apparently bad perturbative convergence raises the question of whether this behaviour is an artifact of the renormalization scale choice. In Fig.~\ref{plt::os} we show the on-shell results and in Fig.~\ref{plt::ms} the $\overline{\text{MS}}$ results, normalized to the LO at $\mu=M_h$.
In both cases, the renormalization scale dependence of the N$^3$LO correction is slightly more stable than at NNLO. In the on-shell scheme the renormalization scale dependence decreases from $0.013\%$ to
$0.009\%$, whereas in the $\overline{\text{MS}}$ scheme the scale dependence decreases from $0.013\%$ to $0.006\%$.
The near-exact cancellation between the NNLO and N$^3$LO corrections in the on-shell scheme only occurs for $\mu \approx M_h$. However, in the whole range of renormalization scales considered,
the N$^3$LO correction is of the same order of magnitude as the NNLO correction, and negative.

It is instructive to study the subsequent terms of the large-mass expansion entering the N$^3$LO
correction to the decay width.
In the case of the logarithm of the gluon-gluon-Higgs form factor
it was found at N$^3$LO, in the on-shell scheme, that the first mass-suppressed term amounts to roughly $10\%$ of the leading term, and the second mass-suppressed term amounts to $1\%$~\cite{Davies:2019wmk}.
This is in contrast to lower orders, where the convergence is better.
However, the gluon-gluon-Higgs form factor is infrared divergent and thus not a physical observable,
unlike the decay width which we consider here.
Normalizing the NNLO and N$^3$LO contributions to $\Gamma_{h\rightarrow \gamma\gamma}^{\text{OS}}$ to their leading term in the large-mass expansion, we obtain for $\mu = M_h$:
\begin{align}
\frac{\Gamma_{h\rightarrow \gamma\gamma}^{\text{OS}}|_{\text{NNLO}}}{\Gamma_{h\rightarrow \gamma\gamma}^{\text{OS}}|_{\text{NNLO},\rho\rightarrow 0}} = 1 - 0.517 - 0.071 - 0.007~,\nonumber\\
\frac{\Gamma_{h\rightarrow \gamma\gamma}^{\text{OS}}|_{\text{N$^3$LO}}}{\Gamma_{h\rightarrow \gamma\gamma}^{\text{OS}}|_{\text{N$^3$LO},\rho\rightarrow 0}} = 1 + 0.560 + 0.084 + 0.010~,
\end{align}
where we have separated the $\rho^0$, $\rho^1$, $\rho^2$ and $\rho^3$ contributions.
Thus, both for NNLO and N$^3$LO the mass-suppressed terms are larger than for the gluon-gluon-Higgs form factor. However, the mass corrections at N$^3$LO are only slightly larger than their NNLO counterparts.
Taking $\mu = \sqrt{2} M_h$ as a renormalization scale, we obtain
\begin{align}
\frac{\Gamma_{h\rightarrow \gamma\gamma}^{\text{OS}}|_{\text{NNLO}}}{\Gamma_{h\rightarrow \gamma\gamma}^{\text{OS}}|_{\text{NNLO},\rho\rightarrow 0}} = 1. - 0.302 - 0.039 - 0.004~,\nonumber\\
\frac{\Gamma_{h\rightarrow \gamma\gamma}^{\text{OS}}|_{\text{N$^3$LO}}}{\Gamma_{h\rightarrow \gamma\gamma}^{\text{OS}}|_{\text{N$^3$LO},\rho\rightarrow 0}} = 1. + 3.686 + 0.529 + 0.060~,
\end{align}
and for $\mu = M_h/\sqrt{2}$ we obtain
\begin{align}
\frac{\Gamma_{h\rightarrow \gamma\gamma}^{\text{OS}}|_{\text{NNLO}}}{\Gamma_{h\rightarrow \gamma\gamma}^{\text{OS}}|_{\text{NNLO},\rho\rightarrow 0}} = 1. + 2.257 + 0.348 + 0.038~,\nonumber\\
\frac{\Gamma_{h\rightarrow \gamma\gamma}^{\text{OS}}|_{\text{N$^3$LO}}}{\Gamma_{h\rightarrow \gamma\gamma}^{\text{OS}}|_{\text{N$^3$LO},\rho\rightarrow 0}} = 1. + 0.274 + 0.043 + 0.005~.
\end{align}
In all of the above cases, the mass-suppressed terms amount to at least $25\%$ of the leading term, and in some cases as much as $400\%$.
However, in all cases, the corrections at order $\rho^3$ are sufficiently small.
Also in all of the above cases, the $\overline{\text{MS}}$ scheme displays better convergence. For brevity we do not show them here.

In the case of the gluon-gluon-Higgs form factor, it was observed that the mass-suppressed terms become
increasingly important at higher perturbative orders. However in this case, no such clear statement can
be made; the convergence of the large-mass expansion at different perturbative orders depends strongly
on the chosen value of the renormalization scale.
The scheme difference decreases from $0.013\%$ to $0.004\%$ for $\mu = M_h$ and the decrease is approximately constant across renormalization scales in the range $(M_h/\sqrt{2},\sqrt{2} M_h)$.

A final remark concerning the NNLO real-radiation corrections, computed in Section~\ref{sec:real},
is in order. They amount to
\begin{align}
\label{eq::nnlo-real-numerics}
\Gamma^{\text{OS}}_{h\rightarrow \gamma\gamma g g} \times 10^{16} \text{GeV}^{-1} &= 1.68 + 0.48 + 0.09 = 2.25~,
\end{align}
where we show the individual terms in the large-mass expansion. They converge well, but are clearly negligible\footnote{While these terms are both mass suppressed and come with small coefficients, it might be
worthwhile to study the effect for bottom quark loops. It is possible that they are larger, despite the $m_b$ suppression.}.

\subsection{\label{sec::all} Including light quarks and electroweak corrections}
We are now in the position to combine the N$^3$LO corrections computed in the previous section with the known NLO electroweak corrections \cite{Actis:2008ts}, as well as the NNLO QCD corrections with
massive bottom and charm quark loops \cite{Niggetiedt:2020sbf}. To this end, we convert the results of \cite{Niggetiedt:2020sbf}, which are given in terms of the on-shell quark mass, into the $\overline{\text{MS}}$ scheme
and transform $\alpha_s$ to the five-flavour scheme. A detailed discussion of the conversion can be found in Appendix \ref{sec::lq}. Note that here the top quark mass is in the on-shell scheme, to be compatible
with the NLO electroweak corrections of \cite{Actis:2008ts}.
Furthermore, we include $\tau$ loops at LO.

For the light fermion masses we take \cite{Zyla:2020zbs}
\begin{align}
m_b^{(5)}\left(m_b\right) = 4.18~\text{GeV},\quad m_c^{(4)}\left(m_c\right) = 1.27~\text{GeV}\quad \text{and} \quad M_\tau = 1.77686~\text{GeV}~.
\end{align}
We employ the program \texttt{RunDec} to obtain the two light quark masses in the five-flavour scheme at $\mu = M_h$.

As result, we obtain
\begin{align}
\Gamma_{h\rightarrow \gamma\gamma} \times 10^{6} \text{GeV}^{-1} &= 9.2581|_{\text{LO}} - 0.1502|_{\text{NLO,EW}} + 0.1569|_{\text{NLO,t}} + 0.0157|_{\text{NLO,bc}}\nonumber\\
&+ 0.0030|_{\text{NNLO,t}} + 0.0037|_{\text{NNLO,bc}} - 0.0031|_{\text{N$^3$LO,t}} = 9.2840~,
\end{align}
where we split the higher-order corrections into electroweak (EW), top-induced (t) and bottom/charm-induced (bc) contributions\footnote{The values for the top quark--induced terms differ from the values in Eq.~(\ref{eq::Gamma-muMh-MtOS}), since in contrast to the previous section we also include the light fermions at LO.}.
We note that the NNLO corrections coming from bottom and charm quark loops are of the same order as the NNLO top quark contributions, as are the N$^3$LO top quark corrections.
This seems to indicate that the NNLO top quark contributions are ``accidentally small'', since the bottom and charm contributions themselves seem to converge well.

\subsection{\label{sec::unc} Theoretical uncertainties}
We can now comment on the sources of theoretical uncertainties on the partial decay width; missing higher-order QCD and electroweak corrections, as well as parametric uncertainties.

The N$^3$LO QCD corrections involving top quark loops and the NNLO QCD corrections involving bottom and charm quark loops are both smaller than $0.05\%$ while the scale and scheme dependence is even smaller.
Furthermore, the mixed quark-flavour
contributions should be well approximated by our treatment (see Appendix~\ref{sec::lq}).
Thus, we assign an overall uncertainty of $0.1\%$ for missing higher-order QCD corrections.

In contrast to the pure QCD corrections, the effects of three-loop electroweak and mixed QCD-electroweak corrections are harder to estimate. While naive arguments, such as the $\mathcal{O}\left(\alpha_s\right)$ suppression
w.r.t.~the NLO corrections would suggest that they are smaller than $1\%$, counter-examples for similar quantities exist in the literature. For example
in the case of the Higgs decay into gluons, the NNLO $\mathcal{O}\left(G_F M_t^2 \alpha_s^3\right)$ corrections
are of the same order of magnitude as the NLO $\mathcal{O}\left(G_F M_t^2 \alpha_s^2\right)$ corrections \cite{Steinhauser:1998cm}. Furthermore, there is a significant cancellation between
fermionic contributions at NLO and purely bosonic contributions\footnote{See Table 2 in \cite{Degrassi:2005mc}.}, which are not necessarily the case at NNLO. As a consequence, we conservatively assign an uncertainty of $1.6\%$
due to missing higher-order electroweak corrections, which is the size of the NLO electroweak corrections.

To obtain the parametric uncertainties, we vary the input parameters within the experimental uncertainties. The dominant parametric uncertainty stems from the Higgs boson mass
and amounts to $\pm 0.5\%$. All other parametric uncertainties are negligible.

\section{Conclusion and outlook}
In this paper we investigate the top quark contributions to the Higgs boson decay into photons at N$^3$LO in QCD and compute real-radiation corrections at NNLO. While the corrections are small and will not play a role in
the precision programme of the LHC, we find that the N$^3$LO corrections are of the same order of magnitude as the NNLO corrections and that the mass-suppressed terms are sizeable relative to
the leading term in the large-mass expansion; they must be included for a reasonable description of the
top quark mass effects.

We combine our results with the recently-obtained NNLO QCD corrections due to bottom and charm loops, as well as NLO electroweak corrections and, for the first time,
quantify the residual uncertainty due to higher-order QCD corrections. We obtain
\begin{align}
\Gamma_{h\rightarrow \gamma\gamma} = 9.284 \pm 0.009|_{\text{QCD}} \pm 0.15|_{\text{EW}} \pm 0.046|_{M_h}~\text{keV}~,
\end{align}
where the three different sources of uncertainties are discussed in Sec.~\ref{sec::unc}. We find that the largest source of theoretical uncertainty on $\Gamma_{h\rightarrow \gamma\gamma}$
are the missing NNLO electroweak and mixed QCD-electroweak corrections, which would be required to reduce the total uncertainty to 1\% and below.

\section*{Acknowledgements}
We thank Robert Harlander for a question leading to this work and Christian Sturm for communication regarding the NLO electroweak corrections. We are grateful to Matthias Steinhauser
for many discussions on the topic, as well as carefully reading the manuscript.

The work of J.D. was in part supported by the Science and Technology Facilities Council (STFC) under the Consolidated Grant ST/T00102X/1.
This document was prepared using the resources of the Fermi National Accelerator Laboratory (Fermilab), a U.S. Department of Energy, Office of Science, HEP User Facility. Fermilab is managed by Fermi Research Alliance, LLC (FRA), acting under Contract No. DE-AC02-07CH11359.
This research was supported by the Deutsche Forschungsgemeinschaft (DFG, German Research Foundation) under grant 396021762 - TRR257 ”Particle Physics Phenomenology after the Higgs Discovery”.

\appendix
\section{\label{sec::lq} Light quark contributions}
In this section we briefly describe the steps necessary to adapt the results of \cite{Niggetiedt:2020sbf} to the conventions used in this paper. The results of \cite{Niggetiedt:2020sbf} are given for $\mu = M_h$, the quark mass
renormalized in the on-shell scheme and $\alpha_s$ in the four-flavour scheme for bottom quark and three-flavour scheme for charm quark. Thus several steps have to be taken to obtain
results for arbitrary $\mu$ and the quark masses renormalized in the $\overline{\text{MS}}$ scheme.

We write the bottom quark contribution to the amplitude in terms of the on-shell quark mass $M_b$ as
\begin{align}
\label{eq::amp-os}
\mathcal{A}_b^{\text{OS}} = \mathcal{A}^{(0)}_b\left(M_b\right) + \frac{\alpha_s^{(4)}\left(M_h\right)}{\pi}\mathcal{A}^{(1)}_b\left(M_b\right) + \left(\frac{\alpha_s^{(4)}\left(M_h\right)}{\pi}\right)^2\mathcal{A}^{(2)}_b\left(M_b\right)~.
\end{align}
We first express $\alpha_s^{(4)}\left(M_h\right)$ in the five-flavour scheme and for an
arbitrary renormalization scale $\mu$,
\begin{align}
\frac{\alpha_s^{(4)}\left(M_h\right)}{\pi} = \frac{\alpha_s^{(5)}\left(\mu\right)}{\pi} + \left(\frac{\alpha_s^{(5)}\left(\mu\right)}{\pi}\right)^2\left(\frac{\beta_0}{4}\ln\left(-\frac{\mu^2}{M_h^2}\right) + \zeta_{\alpha_s}^{(1)}\right)~.
\end{align}
Here $\beta_0 = \frac{11}{3}C_A - \frac{4}{3}T_f n_l$ is the one-loop QCD $\beta$-function and $\zeta_{\alpha_s}^{(1)} = -\frac{T_F}{3}\ln\left(\frac{\mu^2}{M_b^2}\right)$ is the one-loop contribution to the decoupling constant of $\alpha_s$.

Next, we convert the on-shell mass $M_b$ to the $\overline{\text{MS}}$ mass $m_b^{(5)}\left(\mu\right)$.
We proceed by replacing the on-shell mass in the exact LO and NLO expressions in terms of the
$\overline{\text{MS}}$ mass, according to
\begin{align}
M_b = m_b^{(5)}(\mu) \left(
	1
	+ c^{(1)} \frac{\alpha_s^{(5)}\left(\mu\right)}{\pi}
	+ c^{(2)} \left(\frac{\alpha_s^{(5)}\left(\mu\right)}{\pi}\right)^2
\right)\,,
\end{align}
and then expand in $\alpha_s$. Here $c^{(l)}$ are the $l$-loop coefficients of the
OS-$\overline{\text{MS}}$ relation and depend on $\mu$ and $m_b^{(5)}\left(\mu\right)$.
This yields an expression for the $\overline{\text{MS}}$ amplitude in terms of the on-shell amplitude
expansion coefficients of Eq.~(\ref{eq::amp-os}),
\begin{align}
\mathcal{A}_b^{\overline{\text{MS}}} ={}& \mathcal{A}^{(0)}_b\left(m_b^{(5)}\left(\mu\right)\right)
+ \frac{\alpha_s^{(5)}\left(\mu\right)}{\pi}\left(\mathcal{A}^{(1)}_b\left(M_b\right) + c^{(1)}M_b\frac{\partial}{\partial M_b}\mathcal{A}^{(0)}_b\left(M_b\right)\right)\Bigg|_{M_b = m_b^{(5)}\left(\mu\right)}\nonumber\\
&+ \left(\frac{\alpha_s^{(5)}\left(\mu\right)}{\pi}\right)^2\Bigg(\mathcal{A}^{(2)}_b\left(M_b\right) + \left(c^{(1)}M_b\frac{\partial}{\partial M_b} + \frac{\beta_0}{4}\ln\left(-\frac{\mu^2}{M_h^2}\right) +
\zeta_{\alpha_s}^{(1)}\right)\mathcal{A}^{(1)}_b\left(M_b\right)\nonumber\\
&+\left(c^{(2)}M_b\frac{\partial}{\partial M_b} +\frac{\left(c^{(1)}\right)^2}{2}M^2_b\frac{\partial^2}{\partial M_b^2}\right)\mathcal{A}^{(0)}_b\left(M_b\right)\Bigg)\Bigg|_{M_b = m_b^{(5)}\left(\mu\right)}~,
\end{align}
For the NLO contribution only the first line is necessary.
Note that here the charm quark is treated as a massless quark.

For the charm quark we start out in the three-flavour scheme, where in addition to the above steps a decoupling
of the bottom quark in both $\alpha_s$ and $m_c$ has to be performed.

\end{document}